\documentclass[12pt]{iopart}
\usepackage{iopams}  
\usepackage{graphicx}

\begin{document}

\title[Curvature Correction in the Strutinsky's Method]{Curvature Correction in the Strutinsky's Method}

\author{P Salamon$^{1,2}$ and A T Kruppa$^1$}

\address{$^1$Institute of Nuclear Research of the Hungarian Academy of Sciences, 
Debrecen, Bem t{\'e}r 18/c, Hungary H-4026,\\$^2$University of Debrecen\\
Faculty of Informatics,Debrecen, P.O.Box 12, Hungary H-4010}
\ead{salamon@atomki.hu}
\ead{atk@atomki.hu}

\begin{abstract}

Mass calculations carried out by Strutinsky's shell correction method are based on 
the notion of smooth single particle level density. The smoothing procedure is always 
performed using curvature correction. In the presence of curvature correction a smooth function 
remains unchanged if smoothing is applied. Two new curvature correction methods are introduced. 
The performance of the standard and new methods are investigated using harmonic oscillator and 
realistic potentials.  
\end{abstract}
\submitto{\JPG}
\maketitle

\section{Introduction}

Masses, or, equivalently ground state energies of nuclei are very important basic quantities. For example, shell structure changes around drip lines or existence of new magic numbers beyond the heaviest known nuclei can be studied with the help of mass measurements. Important applications are the determination of possible decay modes of a system and the calculation of  energy release in a given reaction.
Mass values are important not only for nuclear physics but for astrophysics and particle physics as well. Nowadays not only the precision of mass measurements has improved dramatically but it is possible to measure masses of short lived nuclei \cite{Lun03,Bla06}. Very accurate experimental mass values are now available for almost every part of the periodic table and this is a great challenge
for theoretical description. 

Global mass calculations have been carried out using the so called macroscopic-microscopic method \cite{Bha09,Mo95}. In this approach 
the binding energy is the sum of two terms
$
E(N,Z)=E_{mac}(N,Z)+E_{mic}(N,Z).
$
The first term  $E_{mac}(N,Z)$ is a smooth function of the neutron  ($N$) and proton number ($Z$). It  can be determined  using a macroscopic model. The simplest macroscopic expression is derived from the liquid drop model. This 
form can be generalized in several respects. E.g. a well known macroscopic term is 
the formula of the finite range droplet model \cite{Mo95}.

Beside the smooth trend the binding energy formula contains an oscillatory part with pure quantum origin.
The microscopic part $E_{mic}(N,Z)$ consists of two terms shell correction and pairing energy 
$
E_{mic}(N,Z)=E_{shell}(N,Z)+E_{pair}(N,Z).
$
The pairing term is usually calculated using BCS or Lipkin-Nogami scheme and the shell correction by the Strutinsky's method \cite{Stru68}. 

The first step in the calculation of shell correction is to take a single particle Hamiltonian. 
The energy $E$ of the ground state is simply sum of the energies of occupied single particle states. 
This energy can be decomposed into a sum of a smooth $\bar E$ and a fluctuating term $\tilde E$,  
$E=\bar E+\tilde E$. It can be argued using the Strutinsky's energy theorem \cite{Bun72} that 
the quantities  $E_{shell}(N,Z)$ and $\tilde E$ are identical.

Recently new facts have emerged concerning the fluctuating part of the energy. 
The shell correction is a rapidly changing function of the particle number. 
Earlier, the average value of $\tilde E$ was expected to be zero. 
Recently it has been shown that the average value of the fluctuating part of 
the energy is non zero and it contributes to the smooth part of the energy 
\cite{Cen06}. The extra smooth trend of the average value of $\tilde E$ can be 
analytically calculated \cite{Cen06,Roc07} 
in case of  simple model potentials. It would be desirable 
to minimize this extra smooth trend of the shell correction. 
A study of the nuclear mass error distribution \cite{Hi04},  
using  results of the Strutinsky type macroscopic-microscopic  
calculation of \cite{Mo95}, has been revealed a regular correlation and it has been  
suggested that refinement of the Strutinsky's method is needed.

The focus of the present paper is how to 
calculate the smooth part of the energy $\bar E$ or equivalently the shell correction. 
The basic concepts are the quantum single particle level density (SPLD) and the smooth SPLD. 
The quantum SPLD is simply sum of Dirac-delta functions centered at the single particle energies. 
The smooth SPLD is free from singularities and it is a smooth function.
It may be determined using  semi-classical methods \cite{Brakonv}. Shell correction 
calculations \cite{Bha09,Naz94} were carried out 
using the smooth semi-classical 
SPLD based on Wigner-Kirkwood expansion. 
Smooth SPLD can be derived using only the single particle energies and a special smoothing method 
called Strutinsky's smoothing  or standard averaging method \cite{Naz94}.
The Strutinsky type \cite{Mo95} and semi-classical \cite{Bha09} 
macroscopic-microscopic models differ on the way how the smooth SPLD is calculated. 
Beside the energy averaging particle number space averagings were developed \cite{Stru75,Ton82,Pom04} 
in order to calculate $\bar E$ and even the combination of averaging spaces was considered \cite{Dia04}.

The Strutinsky's smoothing of quantum SPLD is carried out with a well defined procedure. 
First, a normalized, positive and even function is selected. 
The smoothing means that the convolution of the quantum SPLD and the kernel function is
calculated \cite{Bun72,Bra73}. This simple procedure however is insufficient. 
We have to demand self-consistency of the smoothing process. If the smoothing is applied to a sufficiently smooth function (polynomial) then the result has to be identical with the original function. 
This property is referred to as curvature corrected smoothing. 

In the most standard form of the Strutinsky's method the curvature correction is 
taken into account by multiplying the starting kernel function with an appropriately 
chosen polynomial \cite{Bun72}.
Brack and Pauli have introduced \cite {Bra73} an alternative method to achieve the 
effect of the curvature correction. In this paper we introduce two new 
curvature correction methods.

The organization of the paper is the following.
In section 2 we describe the Strutinsky's shell correction method and than in section 3 we give precise meaning of the curvature correction. Two standard methods  and our new methods for curvature correction are also presented here. Finally at section 4 we compare the methods using harmonic oscillator and Saxon-Woods potentials. Summary is given at section 5.

\section{Strutinsky's shell correction method}

In the Strutinsky's method first an appropriate single particle Hamiltonian is taken. The eigenvalues of bound states  
and the corresponding degeneracies are denoted by $\epsilon_i$ and $d_i$, respectively. For 
simplicity  the $\{\epsilon_i\}$ series is arranged in increasing order. An important 
notion of the method is the single particle level density (SPLD). The so called exact or quantum mechanical SPLD is of the form
\begin{equation}\label{gdef}
g(\epsilon)=\sum_{i=1}^{N_b}d_i\delta (\epsilon-\epsilon_i),
\end{equation} 
where $\delta (\epsilon-\epsilon_i)$ is the Dirac-delta function and the number of bound states is $N_b$.

The central notion of the Strutinsky's method is the smooth SPLD.
It is denoted by $\bar g(\epsilon)$. The smoothed Fermi-level $\bar\lambda$ is obtained from the 
particle number equation 
\begin{equation}
N_\tau=\int_{-\infty}^{\bar\lambda}\bar g(\epsilon)d\epsilon,
\end{equation}
where $N_\tau$ is the neutron or proton number of the considered nucleus since 
the shell correction is separately calculated for neutrons and protons.
The smoothed total energy $\bar E$ can 
be calculated by an integral
\begin{equation}
\bar E=\int_{-\infty}^{\bar\lambda}\epsilon \bar g(\epsilon)d\epsilon.
\end{equation}
The shell correction is defined by the equation 
\begin{equation}
\tilde E=E_{sp}-\bar E,
\end{equation}
where the total single particle energy $E_{sp}$ is taken in the spirit of 
the independent particle shell model
\begin{equation}
E_{sp}=\sum_{i=1}^{N_\tau}\epsilon'_i.
\end{equation}
The new energy series $\{\epsilon'_i\}$ is derived from the series $\{\epsilon_i\}$ by 
repeating the eigenvalues $\epsilon_i$ $d_i$ times.

\subsection{Smoothing with convolution}

Convolution is an ideal tool to smooth an oscillatory function $y(x)$.
We take a bell shaped kernel function $K(x)$ and the smoothed $\bar y(x)$ function is calculated 
by the convolution of $K(x)$ and $y(x)$ 

\begin{equation}
\label{convol}
\overline{y}(x)={1\over\gamma}\int_{-\infty}^{\infty} K\Big({x-x'\over\gamma}\Big) y(x') dx',
\end{equation}
where the positive real number $\gamma$ defines the smoothing width. 
The smooth SPLD can be obtained from  (\ref{convol}) by identifying $y$ with the quantum SPLD $g$. 
The result is
\begin{equation}\label{gdef1}
\bar g(\epsilon)={1\over\gamma}\sum_{i=1}^{N_b} d_iK\Big({\epsilon-\epsilon_i\over\gamma}\Big).
\end{equation}
The particle number equation takes the form
\begin{equation}\label{pint}
N_\tau=\sum_{i=1}^{N_b}d_i\int_{-\infty}^{\nu_i}K(x) dx
\end{equation}
and the smoothed total energy is
\begin{equation}\label{eint}
\bar E=\sum_{i=1}^{N_b}d_i\left(\gamma 
\int_{-\infty}^{\nu_i}xK(x)dx+\epsilon_i\int_{-\infty}^{\nu_i}K(x)dx\right),
\end{equation}
where $\nu_i=(\bar\lambda-\epsilon_i)/\gamma$.

\section{Curvature correction}

We expect that a proper definition of 
smoothing has the following property: the smoothing does not change a smooth function. 
Here we identify smooth functions with polynomials \cite{Stru68,Bun72}. 
We require 
that the mapping  $y(x)\rightarrow \overline{y}(x)$  
is the identity map on the set of polynomials 
of order at most $(2m+1)$ i.e.
\begin{equation}
\label{pcond}
\overline {p}_m(x)=p_m(x)={1\over\gamma}\int_{-\infty}^{\infty} 
K\Big({x-x'\over\gamma}\Big) p_m(x') dx',
\end{equation}
where $p_m(x)$ is an arbitrary  polynomial of order at most $(2m+1)$.
We will say that a kernel $K$ satisfying (\ref{pcond}) 
has curvature correction of order $m$. 
If we want to emphasize this property the kernel will be denoted by $K_m$ instead of $K$.
A kernel with curvature correction of order zero will be called starting kernel.
If we want to stress actual values of the smoothing range $\gamma$ and 
the order of the curvature correction $m$ we will use the notation $\bar g_{\gamma,m}(\epsilon)$ for 
the smooth SPLD.

\subsection{Standard curvature correction methods}

For completeness we review two standard curvature correction methods. 
In the most widely used method a kernel
with curvature correction of order $m$ is searched in the form \cite{Bun72}
\begin{equation}
\label{curvpol}
K_{m}(t)=P_m(t)K_0(t),
\end{equation}
where $P_m(t)$ is a polynomial of order $2m$
\begin{equation}
\label{pol3}
P_m(t)=\sum_{i=0}^m c_i t^{2i}.
\end{equation}
The coefficients $c_i$ are determined by the condition (\ref{pcond}).   
The following set of equations can be gained \cite{Bun72}
\begin{equation}
\sum_{i=0}^m M_{2(i+j)}(K_0) c_{2i} =\delta_{j,0}\ \ \ 0\le j\le m,
\end{equation}
where the moments of an arbitrary kernel $K$ is defined by 
\begin{equation}
\label{moments}
M_i(K)=\int_{-\infty}^{\infty} t^i K(t) dt,\ \ \ i=0,1,2,\ldots .
\end{equation}
This approach is called polynomial curvature correction (PCC).

Brack and Pauli \cite{Bra73} have taken the kernel with curvature correction of order $m$ in the form 
\begin{equation}\label{bpcc}
K_m(t)=\sum_{i=0}^m a_{2i} \frac{d^{2i}}{dt^{2i}}K_0(t).
\end{equation}
We will refer to this method as Brack and Pauli curvature correction (BPCC) method.
The condition (\ref{pcond})  uniquely determines the coefficients $a_{2i}$. 
The value of $a_0$ is 1.
The other expansion coefficients satisfy the following set of equations

\begin{equation}\label{bpcce}
\sum_{j=1}^m B_{i,j}a_{2j}=-b_{2i}\ \ \ (i=1,\ldots m),
\end{equation}
where $B_{i,i}=1$, $B_{i,j}=0\ (i<j)$, $B_{i,j}=b_{2(i-j)}\  (1<i<j)$ 
and $b_{2i}$ is related to the moments of $K_0$
\begin{equation}
\label{bm2}
b_{2i}={M_{2i}(K_0)\over (2i)!}. 
\end{equation}

\subsection{Curvature correction with many widths}

The error term of the smoothing 
\begin{equation}
\delta y(x)=\overline{y}(x)-y(x)
\end{equation}
can be cast into the form 
\begin{equation}
\label{error}
\delta y(x)=\sum_{n=1}^{\infty} (-1)^n\gamma^n{1\over n!} M_n(K) y^{(n)}(x),
\end{equation}
where the $n$-th derivative of $y(x)$is denoted by $y^{(n)}(x)$ and it is assumed $M_0(K)=1$.

If the moments of the kernel function $K$ satisfy the following set of equations  $M_0(K)=1$,
$M_i(K)=0$, $i=1, 2, \dots, (2m+1)$ then
from (\ref{error}) it is clear that such a kernel function
has curvature correction of order $m$.

Let us assume that we have a kernel $K_{m}$ and suppose that it is an even function. We claim that a kernel with curvature correction of order $(m+1)$ can be written in the 
form 
\begin{equation}
\label{recursionc}
K_{m+1}(t)={K_m(t)-c^{-2m-3} K_m\Big({t\over c}\Big)\over 1-c^{-2m-2}},\ \  
m\ge 0.
\end{equation}
Here $c$ is an arbitrary real number but $c\neq 0$ and $c\neq 1$. 

In order to prove our claim it is enough to show the validity of the equations 
\begin{equation}
\label{proof1}
\int_{-\infty}^{\infty} K_{m+1}(t) dt=1 
\end{equation}
and 
\begin{equation}
\label{proof2}
\int_{-\infty}^{\infty} t^i K_{m+1}(t) dt=0,\ \ \ \ i=1, 2, \ldots, (2m+3).
\end{equation}
The proofs are very simple. We have to substitute (\ref{recursionc}) into (\ref{proof1}) and
 (\ref{proof2}), use the linearity of the integration  and in the second term of the left hand sides 
change the integration variable 
from $t$ to $u=t/c$.

If the same $c$ value is used at each iteration step (starting from $m=0$) then the following closed expression 
for the kernel $K_m$ ($m \geq 1$) can be derived 
\begin{equation}
\label{kernelm}
K_m(t)=
A_m\sum_{k=0}^m (-1)^k c^{(m-k)(m-k+2)} \Bigg[{m\atop k}\Bigg]_{c^2} K_0\Bigg({t\over c^k}\Bigg),
\end{equation}
where the notation
\begin{equation}
\label{eq10}
\Bigg[{m\atop r}\Bigg]_q=\left\{
\begin{array}{rl}
1~~~~~~~~~~~~~~~~~ &\textrm{ if } r=0\\
{(1-q^m) (1-q^{m-1}) \dots (1-q^{m-r+1}) \over (1-q) (1-q^2) \dots (1-q^r)} &\textrm{ if } 1\leq r\leq m\\
0~~~~~~~~~~~~~~~~~ &\textrm{ if } r>m\\
\end{array}
\right.
\end{equation}
stands for the Gaussian coefficient (q-binomial coefficient) and 
\begin{equation}
A_m=\Bigg( \sum_{k=0}^m c^{2k} \Bigg) {1\over c^{(m+2)^2-4}} 
\Bigg( \prod_{k=1}^m {1\over 1-c^{-(2k+2)}} \Bigg).
\end{equation}
Kernels derived using (\ref{recursionc}) or (\ref{kernelm}) are called many width kernels and the smoothing
method is named as many width curvature correction (MWCC). The name comes from the form of the smooth SPLD of the MWCC method 
\begin{equation}
\label{gm}
\bar g_{\gamma,m}(\epsilon)=
A_m\sum_{k=0}^m (-1)^k c^{(m-k)(m-k+2)+k} \Bigg[{m\atop k}\Bigg]_{c^2}\bar g_{\gamma c^k,0}(\epsilon).
\end{equation}
The smooth SPLD corresponding to a many width kernel with curvature correction of order $m$ is a linear
combination of smooth SPLD's derived from the starting kernel having smoothing width $\gamma c^k\ \ \ (k=0,\ldots m)$. 
The individual terms $\bar g_{\gamma c^k,0}(\epsilon)$ behave differently, some terms may show the shell structure, other terms may over smooth  but the proper linear combination can make a
balance. 

\subsection{Derivative curvature correction}

The recursion (\ref{recursionc}) is not valid at $c=1$. If the 
limit $c \to 1$ is taken using the L'Hospital rule then a new kernel with curvature correction 
of order $m$ emerges  
\begin{equation}
\label{recursiond}
K_{m+1}(t)={2m+3\over 2m+2} K_m(t) + {1\over 2m+2} t \frac{d}{dt}K_m(t).
\end{equation}
Kernels obtained in this way are called derivative kernels and the method is called derivative curvature
correction (DCC).

With the aid of the recursion (\ref{recursiond}) a closed expression can be derived 
for the
derivative kernels ($m\ge 0$)
\begin{equation}
\label{kerneld}
K_m(t)=\sum_{k=0}^m {a(m+1,k+1)\over m! 2^m} t^k {d^k\over dt^k} K_0(t),
\end{equation}
where the coefficients $a(n,m)$  obey the recursion
\begin{equation}
\label{arecursion}
a(n,m)=(2n+m-2)a(n-1,m)+a(n-1,m-1)
\end{equation}
with initial conditions $a(1,1)=1$, $a(n,0)=0$ ($n$ is arbitrary) and $a(n,m)=0$ for $n<m$. 
Apart from the multiplicative factor $t^k$ the kernel (\ref{kerneld}) is similar to the kernel (\ref{bpcc}). In contrast to the BPCC method the linear combination coefficients 
in (\ref{kerneld}) are independent from the starting kernel. In the BPCC
method the linear combination coefficients are calculated using (\ref{bpcce}).

The recursion (\ref{recursiond}) simplifies the calculation of auxiliary quantities
of the shell correction method.
For the calculation of the particle number we need the integral (see  (\ref{pint}))
\begin{equation}
\label{eq12}
N_m(x)=\int_{-\infty}^x K_m(t)dt,~~~m\geq 0.
\end{equation}
We can easily get the following recursion for the function $N_m(x)$
\begin{equation}
\label{eq13}
N_{m+1}(x)=N_m(x)+{1\over 2m+2} x K_m(x).
\end{equation}
This recursion is valid provided
$\lim_{t \rightarrow -\infty} \Big( t K_m(t) \Big)=0,~~~m\geq 0.$
For the calculation of the smoothed energy we need the integral (see (\ref{eint}))
\begin{equation}
\label{eq15}
E_m(x)=\int_{-\infty}^x t K_m(t) dt,~~~m\geq 0.
\end{equation}
We can derive the following recursion
\begin{equation}
\label{eq16}
E_{m+1}(x)={2m+1\over 2m+2} E_m(x) + {1\over 2m+2} x^2 K_m(x).
\end{equation}
Here we assumed $\lim_{t \rightarrow -\infty} \Big( t^2 K_m(t) \Big)=0,~~~m\geq 0.$

\section{Results}

At the beginning of an application a starting kernel 
has to be chosen. The most widely used form corresponds to a 
Gaussian $K_0(t)=\exp (-t^2)/\sqrt {\pi}$. 
The kernel $K_0(t)=1/(2 \textrm{ Cosh}^2(t))$ was considered 
in \cite{Bra73} using the BPCC method. We will use both functions and call them Gaussian 
and Cosh-type kernels respectively. An interesting observation have to be mentioned. In the 
case of Gaussian starting kernel the methods PCC, BPCC and DCC lead to the 
same kernel $K_m$. The MWCC method however furnishes new kernel function. 
In the case of Cosh-type starting kernel each presented method is different.

The MWCC method in addition to the smoothing width 
$\gamma$ and the order of the curvature correction $m$ contains a new dimensionless 
technical parameter $c$. 
At each fixed values of $\gamma$ and $m$ the value of $c$  is determined with the following procedure. An averaging region around the
investigated particle number $N_\tau$ is selected. In the following calculations the size of the averaging window was 100. Shell corrections of all nucleus
belonging to the averaging window are calculated. The $c$ parameter is used to minimize
the average value of the shell corrections. This procedure is inspired by recent works
\cite{Cen06,Roc07}. The shell correction is a rapidly changing function of the particle number. 
The average value of $\tilde E$ was expected to be zero but it turned out that 
the average of the fluctuating part of 
the energy is non zero and it contributes to the smooth part of the energy 
\cite{Cen06}. 
The value of $c$ is determined in such a way to minimize this extra smooth trend of the
shell correction. 
   
The shell correction method contains two technical 
parameters the smoothing width $\gamma$ and the order of 
the curvature correction $m$. The result is supposed to be independent from the values of 
these parameters if they are selected in a reasonable way. In the case of harmonic 
oscillator or similar potentials (infinite number of bound states) it was 
demonstrated \cite{Bra73} that there is a large region of the smoothing width where the 
shell correction is practically  constant (plateau condition). Therefore the new 
curvature correction methods have to be checked in the case of harmonic 
oscillator potential to see if 
the plateau is remained or not.

The parameter of the harmonic oscillator 
potential ($\hbar\omega=$6 MeV), the considered particle number ($N_\tau$=70) and 
the order of the curvature correction ($m=3$) were chosen to be the same 
as in \cite{Bra73}. In this work the PCC (Gaussian type kernel) and 
the BPCC (Cosh-type kernel) methods were used. 
The shell correction as a function of the 
smoothing width is displayed in Figure \ref{fig1} using Gaussian starting kernel. 
Since the shell spacings are $\hbar\omega$ the shell corrections are calculated 
only for $\gamma>\hbar\omega$ otherwise the effect of shells can be noticed
in the smooth SPLD \cite{Azi06}.

The curve signed by PCC $m=3$ in Figure \ref{fig1} agrees with the result of \cite{Bra73}. 
In order to see the dependence on the order of the curvature correction 
the calculation is repeated for $m=5$ and it is displayed also in Figure 1. 
On the considered $\gamma$ range the plateau region of the MWCC method 
is roughly two times larger than the plateau of the PCC method. Furthermore the 
increase of the order of the curvature correction from 3 to 5 
does not modify the result of the MWCC method but 
there is large change in the case of the PCC method. 

In the case of Cosh-type starting kernel the shell correction is shown in Figure \ref{fig2}. 
Using this kernel the quality  of the DCC, MWCC and BPCC curvature correction 
methods can be judged. In this case also the MWCC method is proved to  
the best, it gives the largest plateau region. The results of the BPCC and  MWCC
methods differ only at large $\gamma$ values. In this large $\gamma$ region only the MWCC 
method keeps the plateau property.
The importance of the large plateau region derives from the fact 
that in global mass calculation for a given nucleus only one 
$\gamma$ value is considered.
It would be desirable to have this fixed $\gamma$ 
value in the plateau region. 

It remains to study the properties of the curvature correction methods in 
the case of realistic potential.  
We restrict 
ourselves for spherically symmetric mean field.
We will consider Saxon-Woods potential with spin-orbit term. 
The geometrical and strength parameters of the Saxon-Woods potential are taken from  
the parametrization given in \cite{Dud81}. There is a great difference between 
harmonic oscillator and realistic potentials. The first one has only bound states whereas the 
Saxon-Woods potential supports  both bound and scattering (continuum) states. 

It is known that the presence of continuum states and the 
finiteness of the number of bound states cause difficulties in 
the shell correction method \cite{Naz94,Ros72}. The formula (\ref{gdef}) for the 
quantum SPLD contains only the contribution of bound states. 
The exact form of the continuum SPLD is known \cite{Ros72} and it was used in 
\cite{Ros72,Ve98}.
The continuum SPLD is expressed by the help of the scattering phase shift. 
In shell correction calculations 
the bound states usually are obtained by diagonalizing the 
Hamiltonian on a given square integrable basis.
On such a basis however it is difficult to calculate the 
continuum SPLD. An elegant way to solve this problem was 
developed in \cite{Kru98} and was used in shell correction calculation \cite{Ve00}.
In this method two diagonalizations have to be carried out. The first diagonalization is 
the standard one (we take the Hamiltonian as it is). In the 
second diagonalization the interaction is switched off 
and only the kinetic energy is diagonalized using the 
same basis as in the first diagonalization. The
results of the second diagonalization are denoted by $\epsilon_i^{(0)}$ and 
the corresponding degeneracies by
$d_i^{(0)}$. The smooth SPLD which contains the effect 
of scattering states reads

\begin{equation}\label{gdef2}
\bar g(\epsilon)={1\over\gamma}\sum_{i} \left(d_iK\Big({\epsilon-\epsilon_i\over\gamma}\Big)
-d_i^{(0)}K\Big({\epsilon-\epsilon_i^{(0)}\over\gamma}\Big)
\right),
\end{equation}
where the summation over $i$ extends to all states gained 
from the diagonalizations.

The value of the parameter $c$ is determined in the same way as in the harmonic oscillator 
case. Since now we have only finite number of bound states the upper bound of the 
averaging window has a natural value $\sum_{i=1}^{N_b} d_i$. It corresponds to a 
nucleus where all bound states are occupied. The lower bound of the averaging window was 
$N_\tau=50$. We have checked that the result is hardly depends on this value.

The neutron shell correction for $^{146}$Gd is displayed in Figures 3 and 4. 
In the case of Gaussian starting kernel (Figure 3) the order of the curvature correction was 
three and five. Results using Cosh-type starting kernel are given in Figure 4 here the order of 
the curvature correction was seven. From Figures 3 and 4 the same conclusion can be drawn. 
In the case of the method MWCC a very large plateau region is present 
(the shell correction is practically independent from the value of $\gamma$). 
The other shell correction methods have no plateau, 
they satisfy only the local plateau condition 
(the derivative of the shell correction with respect to $\gamma$ is zero at some $\gamma$ point). 
The plateau property can be made quantitative if the total variation of the shell correction is determined
in a given $\gamma$ region. Considering Figure 3 the total variation are 0.3 MeV and 3 MeV 
for the MWCC and the PCC methods. In the case of the harmonic oscillator the total variation of the shell
correction is only 0.01 MeV for the MWCC method. The plateau property in the realistic potential case 
is not as good as in the harmonic oscillator case.

The dependence on the order of the curvature corrections is the same as in the harmonic oscillator case. 
The results of the MWCC method is practically independent from the parameter $m$. 
In the case of the methods PCC, DCC and BPCC the shell correction is sensitive to $m$.

\section{Summary}

In the Strutinsky's shell correction method the smooth single 
particle level density is calculated by a convolution.  The aim of 
the curvature correction is to satisfy a self consistency 
condition, the smoothing (convolution) must leave  
a smooth function unchanged. We are aware only two curvature correction methods but 
in global mass calculations only one of them is used. We call the standard method polynomial curvature correction 
since the curvature correction is achieved by multiplying the kernel of the 
convolution by a polynomial. 
Two new curvature correction methods are introduced. 
One of the new methods (we call it many width curvature correction (MWCC)) is 
proved to be the best among the investigated curvature correction methods. 

For potentials without scattering states (harmonic oscillator and similar potentials) the shell 
correction is 
practically constant (plateau condition) on a very large region over the 
smoothing width in each curvature correction methods. 
However the length of the plateau in the MWCC method is much larger 
than in the other methods. 

For potentials supporting scattering states (realistic potentials) the advantage of the MWCC method is more pronounced. 
In the case of each methods except the MWCC method only the local plateau condition is satisfied. 
Due to the lack of convergence in the order of the curvature correction 
it is difficult to define a unique value for the shell correction.
If the calculation is carried out by the  
MWCC method a large plateau  region exists even in the case of realistic potentials and 
the shell correction is quite stable with respect to the increase of the order of 
the curvature correction. 
It seems to us that the MWCC method is better suited for 
shell correction calculation than the standard procedures. 

\ack 
This work has been supported by the Hungarian OTKA fund No. K72357.

\Figures

~

~

\begin{center}
\includegraphics[scale=0.5]{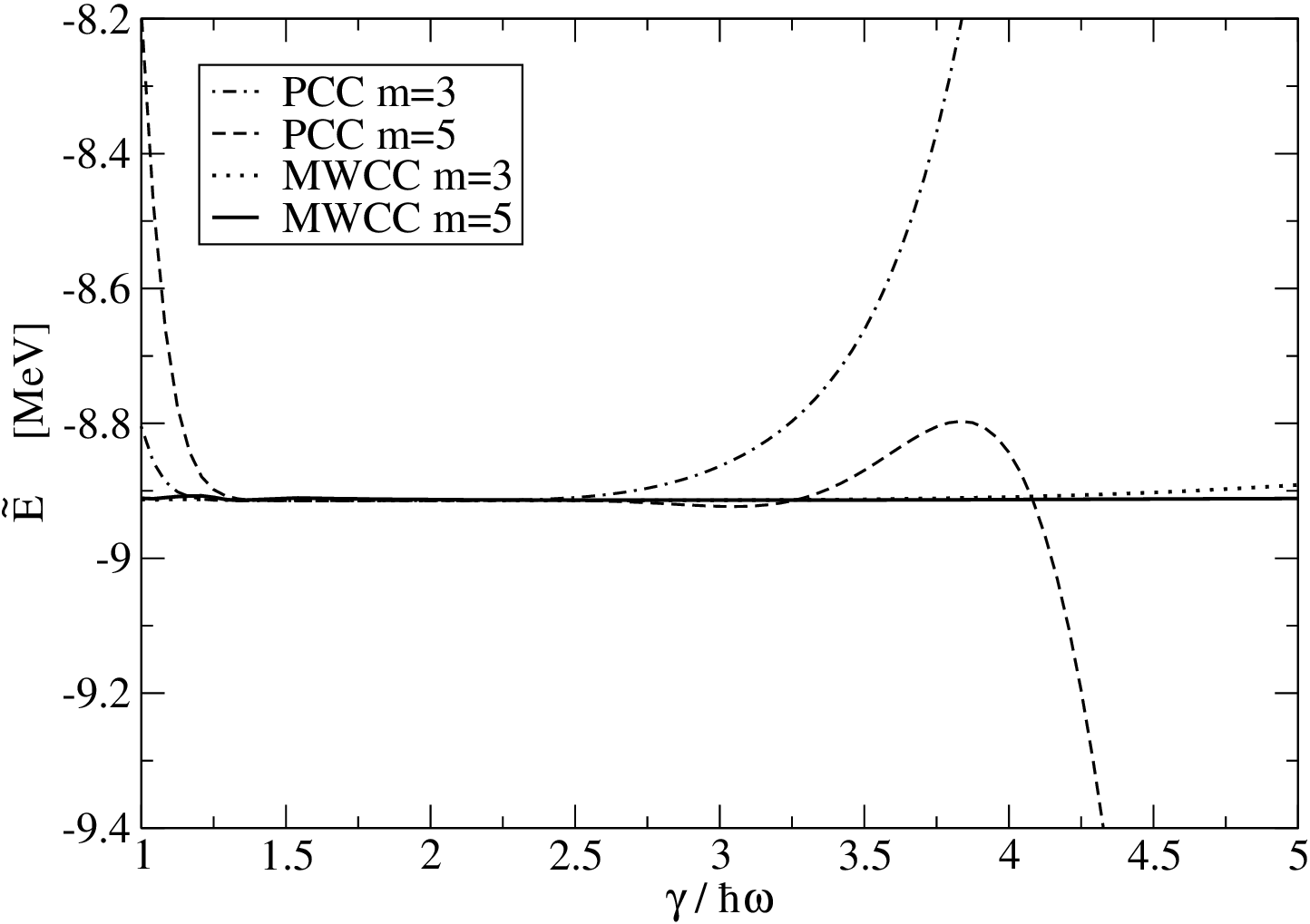}
\end{center}
\Figure{\label{fig1}Shell correction  for the harmonic oscillator potential ($\hbar\omega=6$ MeV, $N_\tau$=70) as a function of the smoothing width using different curvature correction methods. The order of the curvature correction is three and five. The kernels are derived from Gaussian starting kernel. In this case the methods PCC, BPCC and DCC give the same kernel for the smoothing.}
\pagebreak

~

~

\begin{center}
\includegraphics[scale=0.5]{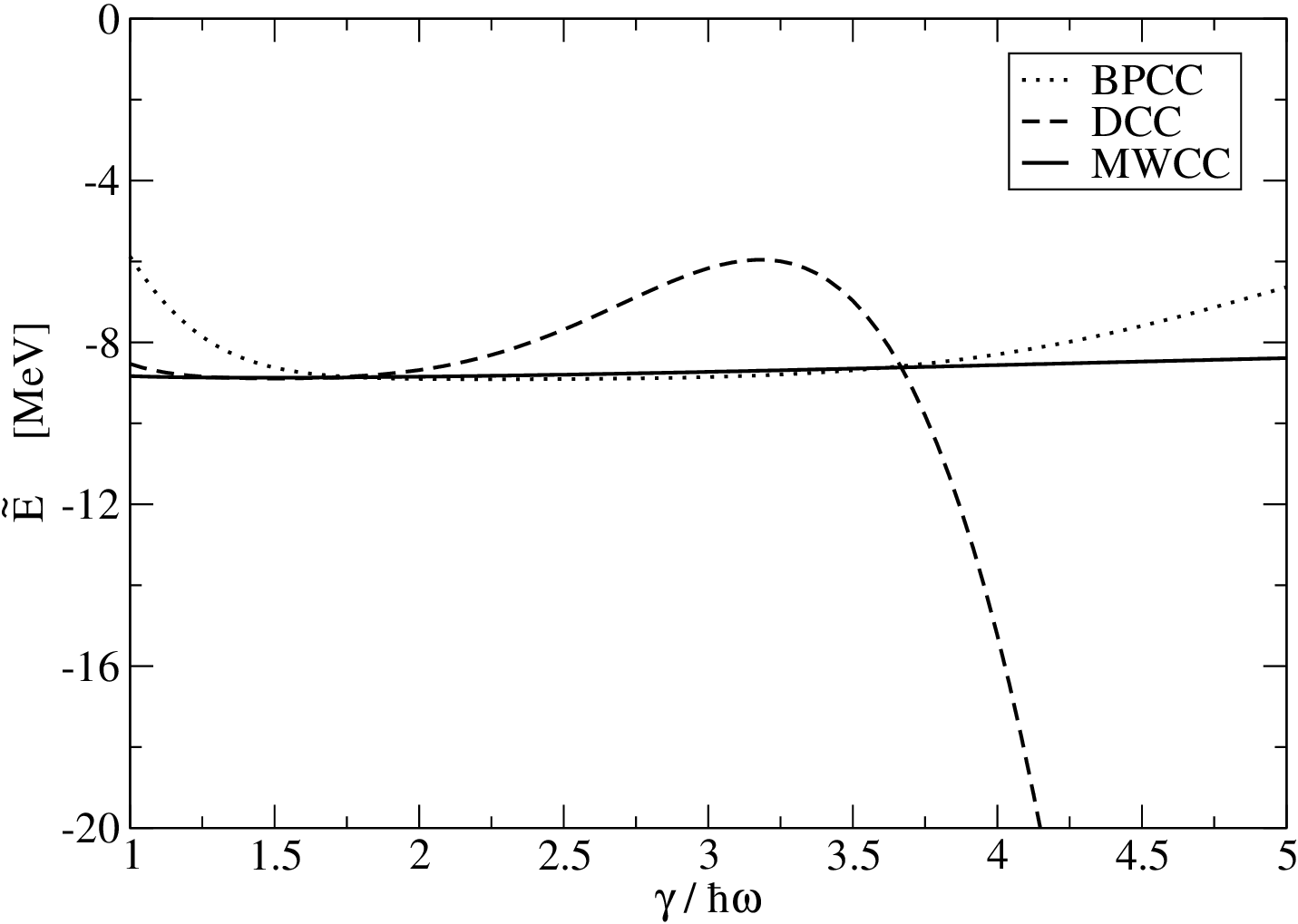}
\end{center}
\Figure{\label{fig2}Same as Figure 1 but the kernels are derived from Cosh-type starting kernel. The order of the curvature correction is three.}
\pagebreak

~

~

\begin{center}
\includegraphics[scale=0.5]{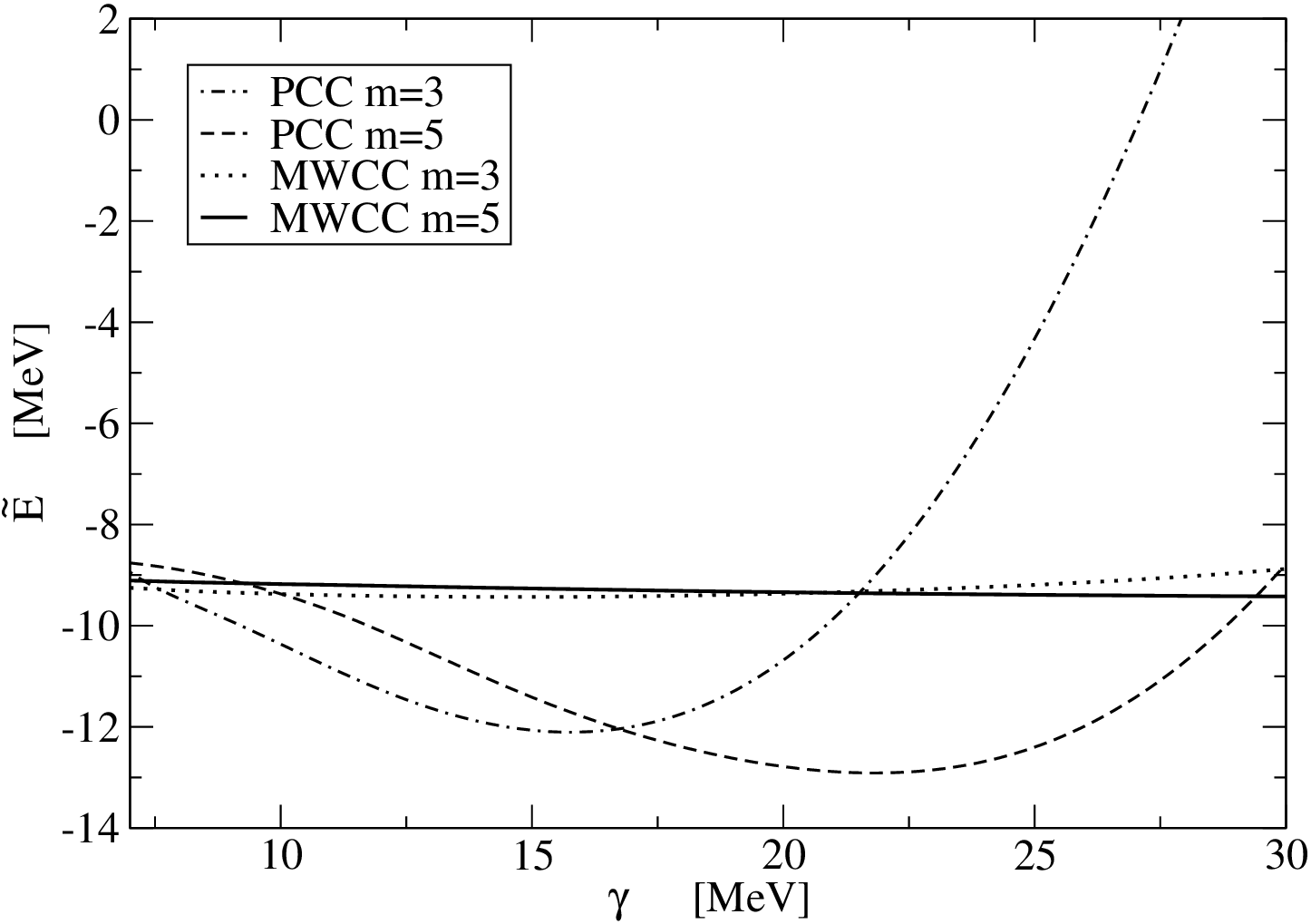}
\end{center}
\Figure{\label{fig3}Neutron shell correction for $^{146}$Gd  as a function of the smoothing width using different curvature correction methods. The single particle mean field was a Saxon-Woods potential with spin-orbit term. The order of the curvature correction is three and five. The kernels are derived from Gaussian starting kernel. In this case the methods PCC, BPCC and DCC give the same kernel for the smoothing.}
\pagebreak

~

~

~

~

\begin{center}
\includegraphics[scale=0.5]{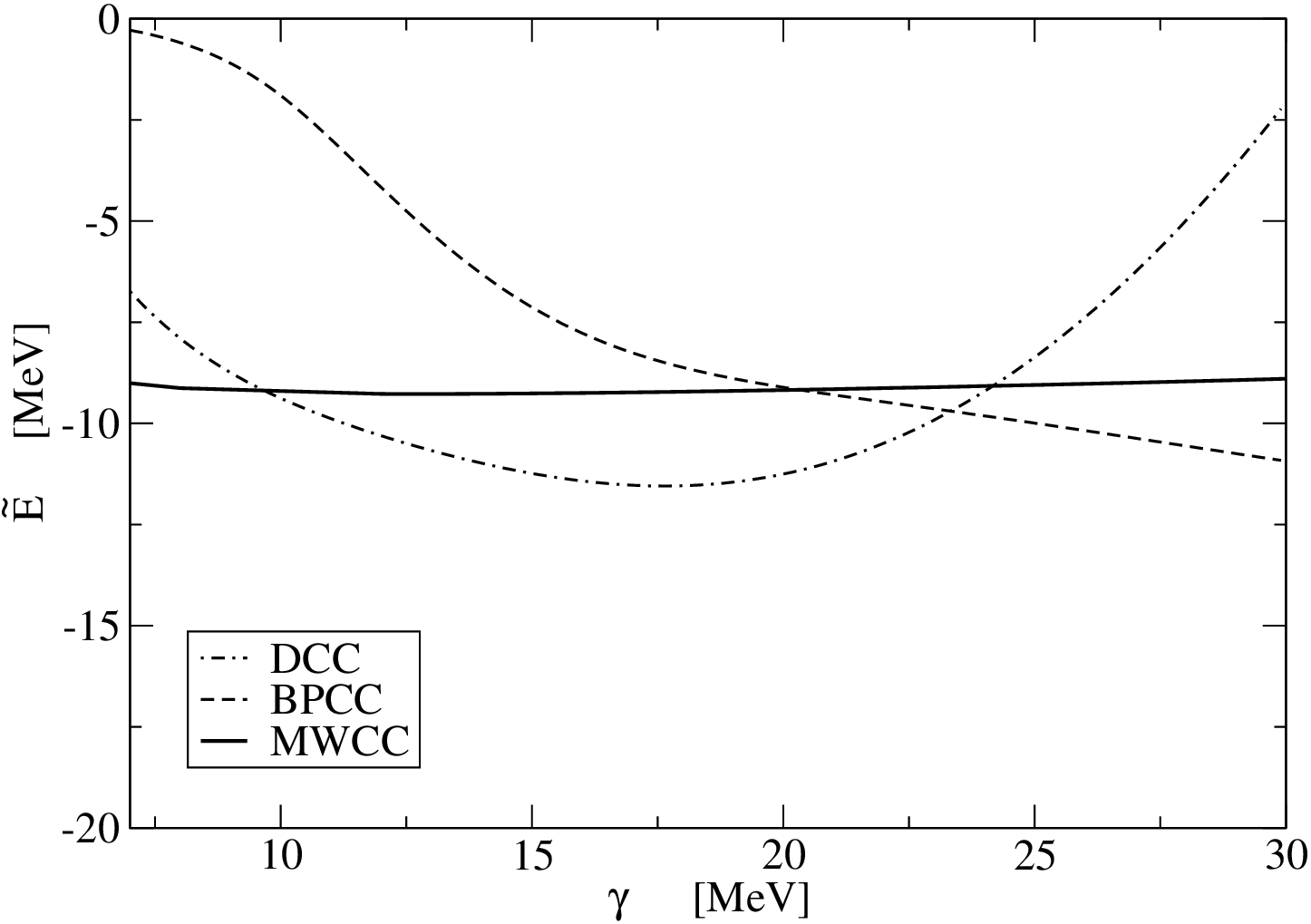}
\end{center}
\Figure{\label{fig4}
Same as Figure 3 but the kernels are derived from Cosh-type starting kernel. The order of the curvature correction is seven.
}

\end{document}